\documentclass[preprint,aps]{revtex4}

\usepackage{graphicx}% Include figure files
\usepackage{dcolumn}% Align table columns on decimal point
\usepackage{bm}% bold math

\newcommand{\be}{\begin{equation}}
\newcommand{\ee}{\end{equation}}
\newcommand{\bey}{\begin{eqnarray}}
\newcommand{\eey}{\end{eqnarray}}
\newcommand{\bw}{\begin{widetext}}
\newcommand{\ew}{\end{widetext}}

\begin{document}

\title{Braid Matrices and Quantum Gates for Ising Anyons
       Topological Quantum Computation}

\author{Zheyong Fan$^{1}$\footnote{\emph{brucenju@gmail.com}}, and
        Hugo de Garis$^{2}$\footnote{\emph{profhugodegaris@yahoo.com}}}
\address{
$^{1}$Department of Physics, Nanjing University,
                Nanjing 210008, China\\
$^{2}$Artificial Intelligence Institute,
       Computer Science Department,\\
       Xiamen University, Xiamen, Fujian Province, China\\}
\date{\today}

\begin{abstract}
We study various aspects of the topological quantum computation scheme based on the non-Abelian anyons corresponding to fractional quantum hall effect states at filling fraction 5/2  using the Temperley-Lieb recoupling theory. Unitary braiding matrices are obtained by a normalization of the degenerate ground states of a system of anyons, which is equivalent to a modification of the definition of the 3-vertices in the Temperley-Lieb recoupling theory as proposed by Kauffman and Lomonaco.  With the braid matrices available, we discuss the problems of encoding of qubit states and construction of quantum gates from the elementary braiding operation matrices for the Ising anyons model. In the encoding scheme where 2 qubits are represented by 8 Ising anyons, we give an alternative proof of the no-entanglement theorem given by Bravyi and compare it to the case of Fibonacci anyons model. In the encoding scheme where 2 qubits are represented by 6 Ising anyons, we construct a set of quantum gates which is equivalent to the construction of Georgiev.
\end{abstract}

\maketitle
\section{Introduction}
\label{intro}
Quantum computers are expected to be able to perform calculations
which are impossible for classical computers, due to quantum
entanglement and  quantum parallelism \cite{nielsen00}.
Unfortunately, quantum computers seem to be extremely difficult to
build because of the unavoidable noise and decoherence caused by
the coupling of the qubits and the ambient environment. It is a
daunting task to construct a quantum computer that has a large
number of qubits and has a low error rate. However, there is a
promising approach, called topological quantum computation (TQC)
\cite{kitaev03,ogburn99,freedman02a,freedman02b,freedman03,dennis02,mochon03,mochon04,kauffman04,preskill219,nayak08,brennen08},
proposing to encode the qubit information into a topological
quantum field. Kitaev \cite{kitaev03} proposed that a system of
anyons can be considered to be a quantum computer. Unitary
matrices are related to moving the anyons around each other.
Measurements are performed by joining anyons in pairs and
observing the result of fusion. Interference experiments
\cite{chung06,stern06,bonderson06a,bonderson06b,feldman06,feldman07} are also proposed to initialize and read out quantum states. The computation is
fault-tolerant by the topological nature of the system.

Different from fermions and bosons, which are the totally
antisymmetric and the symmetric representations of the permutation
group $S_n$, anyons carry fractional charges and have fractional
statistics \cite{wilczek90,frohlich90} which result in nontrivial
phases (for Abelian anyons) or matrices (for non-Abelian anyons)
for permutations. In fact, the underlying symmetry of the system
of anyons is the braid group $B_n$. Abelian anyons correspond to
one-dimensional representations of $B_n$ and the quantum gates one
can construct from them are very limited \cite{averin02}.
Non-Abelian anyons, on the contrary, are much more useful to TQC,
since the braiding of non-Abelian anyons induces non-commuting
(non-Abelian) representations of $B_n$, from which one can
construct various quantum gates.

Physically, anyons  are collective excitations in some condensed
matter systems, such as the fractional quantum hall effect (FQHE)
states  of two dimensional electron liquids. For example, the
effective theories of FQHE states with filling levels $\nu=1/3$, $\nu=5/2$ and $\nu=12/5$ correspond to  Abelian anyons, non-Abelian Moore-Read \cite{moore91} and Read-Rezayi \cite{read99} anyons respectively. Mathematically,  properties of  anyons can be described by $SU(2)_k$ Chern-Simons effective field theories \cite{witten89} and $Z_k$ parafermion conformal field theories (CFT) \cite{zamo85},  the $k=2$ and $k=3$ cases corresponding to the $\nu=5/2$ and the $\nu=12/5$ FQHE states respectively.

The $SU(2)_k$ Chern-Simons theory is a topological quantum field theory \cite{witten89} which  has a deep relationship to knot
invariants, Jones Polynomial \cite{jones85} especially. Kauffman and Lomonaco \cite{kauffman07,kauffman08} studied
unitary representations of braid groups in terms of $q$-deformed
spin networks, or Temperley-Lieb recoupling theory
\cite{kauffman94}.

It is one of our purpose to apply the method of Kauffman and Lomonaco to calculate explicitly the elementary braiding operation (EBO) matrices which govern the exchanges
of Ising anyons, the first non-Abelian anyons model  proposed by Moore and Read \cite{moore91} by constructing a wave function (the Pfaffian
state) for the $\nu=5/2$ FQHE state corresponding to the
$SU(2)_2$ Chern-Simons theory. Direct experimental observation of
fractional electron charge $e/4$ \cite{dolev08,radu08} at the
$\nu=5/2$ FQHE state gives some evidences in support of the
non-Abelian nature of this state. There are many works concerning
the braiding properties of the Ising anyons using CFT method
\cite{nayak96,georgiev09} or quantum group method
\cite{slingerland01}. Nayak and Wilczek \cite{nayak96} suggested
that the Pfaffian wave functions of $n$ ($n$ even) Ising anyons
form a $2^{n/2-1}$ dimensional spinor irreducible representation
of the rotation group $SO(n)$, to which a rigorous treatment is
given by Georgiev recently \cite{georgiev09}. Quantum group
approach \cite{slingerland01} also gives equivalent results. As we
will see, the EBO matrices for
the Ising anyons can be elegantly derived by using the
Temperley-Lieb recoupling theory.

One of the attractive properties of the Ising
anyons TQC model is that the excitation gap at the corresponding
filling fraction $\nu = 5/2$ is the highest one among
all non-Abelian FQHE states, resulting in a very low (or even lower) error rate of $10^{-30}$ \cite{eisenstein02,xia04,sarma05}. Although this Ising anyons model is not universal
\cite{freedman02a,freedman02b,freedman03} for TQC, i.e., the braid group representations are not dense in unitary groups, it receives extensive attention in the past few years
\cite{sarma05,bravyi06,freedman06,georgiev06,georgiev08a,zilberberg08,georgiev08b,ahlbrecht09}.
In fact, it is proved by Bravyi \cite{bravyi06} that no entangled states in the computational space can be obtained purely topologically and the Ising anyons TQC model is classically simulatable. We show that the same conclusion can be obtained from the Temperley-Lieb recoupling theory approach.

We should stress that this no-entanglement theorem does not mean that there is no entanglement between Ising anyons at all. This rule only applies to the qubit encoding scheme (which is consistent with the quantum circuit model) where each qubit is encoded in 4 Ising anyons. Entangled quantum gates, such as controlled-Z and controlled-NOT (CNOT), can be realized purely by braiding 6 Ising anyons with definite topological charge (quantum spin). Using the EBO matrices obtained in this paper, we construct a set of useful 1-qubit and  2-qubit quantum gates, which is not the same as, but equivalent to the construction by Georgiev \cite{georgiev06,georgiev08a}.

The outline of the remainder of this paper is as follows. Section
2 reviews the general models of non-Abelian anyons for TQC and the
formalism of the Temperley-Lieb recoupling theory needed later.
Unitary representations of the braid group in the Hilbert space of
the degenerate ground states of non-Abelian anyons are obtained by
a physical argument which requires that the fusion paths of the
anyons form an orthonomal basis of the Hilbert space. In section 3, we derive the EBO matrices of the Ising anyons model. In section 4, we study some aspects of Ising anyons TQC using the results of section 3. Conclusions and discussions are presented in Section 5.

\section{Temperley-Lieb recoupling theory and
unitary representations of braid groups}
\label{sec:1}
In this section, we first review the definition of the
quantum states of a system of anyons and  the Temperley-Lieb recoupling theory \cite{kauffman94} and then discuss the method to produce unitary representations of  braid groups.

\subsection{Models of non-Abelian anyons for TQC}
\label{sec:2}
A model of non-Abelian anyons consists of the following three
elements \cite{preskill219}: a list of particle types, the fusion
rules, and the braiding rules. In the formalism of $SU(2)_k$
Chern-Simons theory, anyons are quasi-particles having
half-integer $q$-spins (spins for short) $s=0,1/2,1,\cdots,k/2$
as their quantum numbers. The fusion rules of these particles are
truncated versions of the rules of addition of ordinary angular
momenta,
\begin{equation}
s_1 \otimes s_2= |s_1-s_2| \oplus |s_1-s_2|+1 \oplus \cdots \oplus
\textmd{min} (s_1+s_2, k-s_1-s_2).
\label{eq:fusion}
\end{equation}
When a number of non-Abelian anyons with definite spins fuse
consecutively into a single anyon with some spin, the sequences of
the intermediate spins of the fusion paths represent different quantum states of the Hilbert space.

Anyons  commonly appear as collective excitations in 2 dimensional systems. When they move, their world lines propagate in a 3 dimensional space-time. Thus
the exchange of a pair of anyons corresponds to the braiding of
their world-lines. (We will call the braiding of the world-lines
of anyons shortly as the braiding of anyons, but it is important
to keep in mind what it means actually.) In TQC, we perform
quantum computations by braiding anyons to realize certain quantum
gates. Any braiding can be expressed as a sequence of EBOs whose
representations in the above Hilbert space are the EBO matrices we
want to find. The essential task of deriving the EBO matrices is the
determination of the so called R-matrix and F-matrix
introduced first in the context of CFT \cite{moore89}. The former
is the unitary matrix inducing the exchange of neighboring anyons
with definite total spin, and the latter accounts for the
associativity of fusions of anyons. In the next subsection, we
will give their diagrammatic definitions in terms of the
Temperley-Lieb recoupling theory.

\subsection{Temperley-Lieb recoupling theory}

Temperley-Lieb recoupling theory \cite{kauffman94} is based on the Kauffman bracket polynomial model  for the Jones polynomial
at roots of unity and the tangle-theoretic Temperley-Lieb algebra.

\subsubsection{Braid group and Temperley-Lieb algebra}

The Artin braid group  $B_n$ can be presented as a
set of generators $\sigma_1,\sigma_2,...,\sigma_{n-1}$ that obey
the following relations,
\begin{equation}
\begin{array}{l}
\sigma_i\sigma_j=\sigma_j\sigma_i \quad \textmd{for}\quad
|i-j|\geq2;\\
\sigma_i\sigma_{i+1}\sigma_i=\sigma_{i+1}\sigma_i\sigma_{i+1}
\quad \textmd{for}\quad i=1,2,...,n-2.
\end{array}
\label{eq:artin}
\end{equation}
The Temperley-Lieb algebra  $TL_n$ can be presented similarly as a
set of generators $U_1,U_2,...,U_{n-1}$, whose representations are related to the representations $\rho(\sigma_i)$ of $B_n$ by
\begin{equation}
\rho(\sigma_i)= A + A^{-1} U_i,
\label{eq:temperleylieb}
\end{equation}
where the Kauffman variable $A$ is taken to be $A=ie^{i\pi/2r}$
for Jones polynomial at $4r$-th roots of unity such
that the quantum dimension of the spin 1/2 anyon is
$d=-A^{2}-A^{-2}=2\cos(\pi/r)$.

\subsubsection{R-matrix and F-matrix}
\begin{figure}
\resizebox{1.00\columnwidth}{!}{%
  \includegraphics{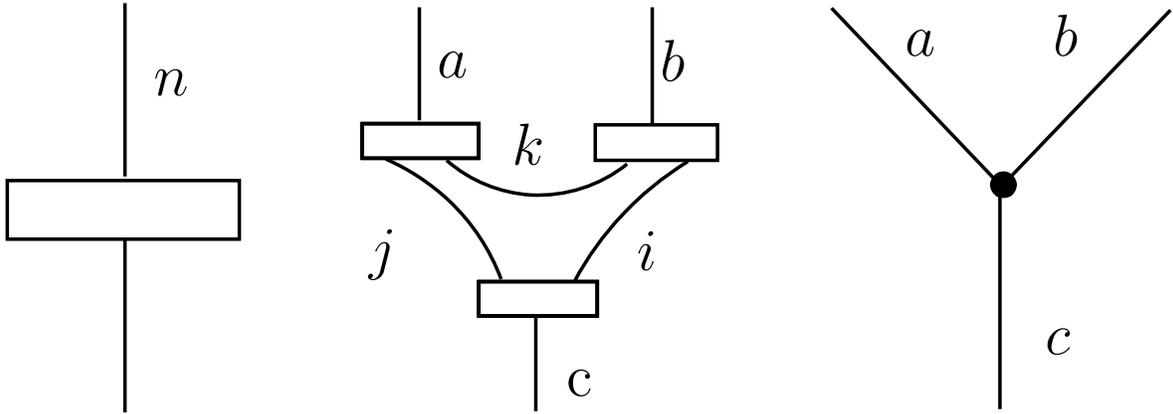}
}
\caption{The Jones-Wenzl projector and the 3-vertex. The integers
$i$, $j$, and $k$ are determined by the relations $a=j+k$,
$b=k+i$, and $c=i+j$.}
\label{fig:jones}       % Give a unique label
\end{figure}

\begin{figure}
\resizebox{1.00\columnwidth}{!}{%
  \includegraphics{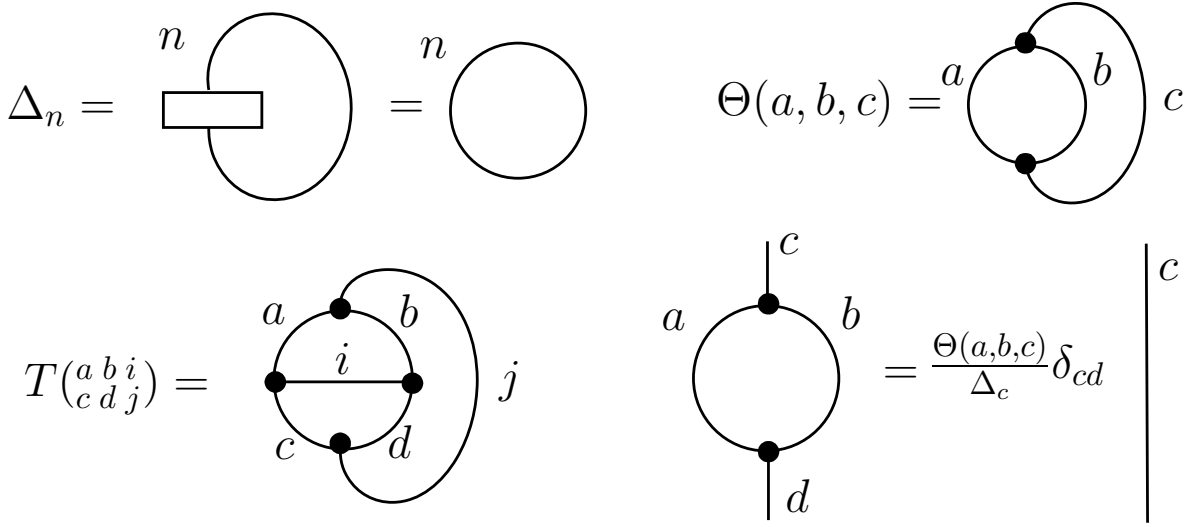}
}
\caption{The definitions of the delta net $\Delta_n$, the theta
net $\Theta(a,b,c)$, and the tetrahedron net
$T(^{a\;b\;i}_{c\;d\;j})$. Formulae \cite{kauffman94} for evaluating these spin-nets are presented in the appendix.}
\label{fig:net}
\end{figure}

The basic object of the Temperley-Lieb recoupling theory is the
Jones-Wenzl projector \cite{kauffman94}. The left graph in Fig. ~\ref{fig:jones}
shows the Jones-Wenzl projector constructed on the basis of the
Kauffman bracket polynomial expansion. The $n$-strand projector
corresponds to the world line of an anyon with spin $n/2$. The
middle graph in Fig. ~\ref{fig:jones} shows the 3-vertex constructed from the
projectors which corresponds to the interaction (fusing or
splitting) of 3 anyons with spins $a/2$, $b/2$, and $c/2$.  The
right graph in Fig. ~\ref{fig:jones} is a simplified notation for the 3-vertex. Note that the $q$-admissible conditions
\cite{kauffman94} for the 3-vertex,
\begin{equation}
\begin{array}{lll}
a+b+c=\textmd{even};\\
a+b-c\geq0,\;b+c-a\geq0,\;c+a-b\geq0; \\
a+b+c\leq2r-4\\
\end{array}
\end{equation}
say exactly the same thing as the fusion rules Eq.~(\ref{eq:fusion}) do due to
the relation $r=k+2$ between $SU(2)_k$ Chern-Simon theory at level
$k$ and Jones polynomial at $4r$-th roots of unity \cite{witten89}
and the fact that the projector with label $n$ represents the
world line of an anyon with spin $n/2$. Various spin networks can
be constructed from Jones-Wenzl projectors and 3-vertices. See
Fig. ~\ref{fig:net}.

\begin{figure}
\resizebox{1.00\columnwidth}{!}{%
  \includegraphics{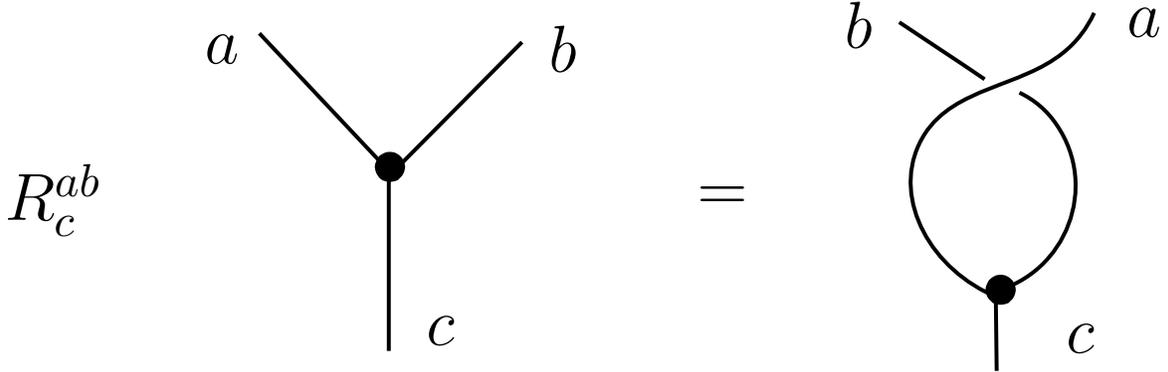}
}
\caption{The action of the R-matrix, whose matrix elements are
given by Eq. ~(\ref{eq:rmatrix}). Our convention is that seen from above, the
braiding of $a$ and $b$ is counterclockwise. }
\label{fig:rmatrix}
\end{figure}

As the world lines of anyons and their fusions being identified
with projectors and 3-vertices, we now consider the braiding
properties of anyons in the context of the Temperley-Lieb
recoupling theory. Fig. ~\ref{fig:rmatrix} shows the braiding of two anyons with
spins $a/2$ and $b/2$ fusing into a spin $c/2$ anyon. Since this
operation dose not change the total spin of the two fusing anyons,
the corresponding matrix, the R-matrix is diagonal in the
underlying Hilbert space. The matrix element $R^{ab}_c$ is given
by the following formula \cite{kauffman94},
\begin{equation}
R^{ab}_c= (-1)^{(a+b-c)/2} A^{-(a(a+2) + b(b+2) - c(c+2))/2}.
\label{eq:rmatrix}
\end{equation}

\begin{figure}
\resizebox{1.00\columnwidth}{!}{%
  \includegraphics{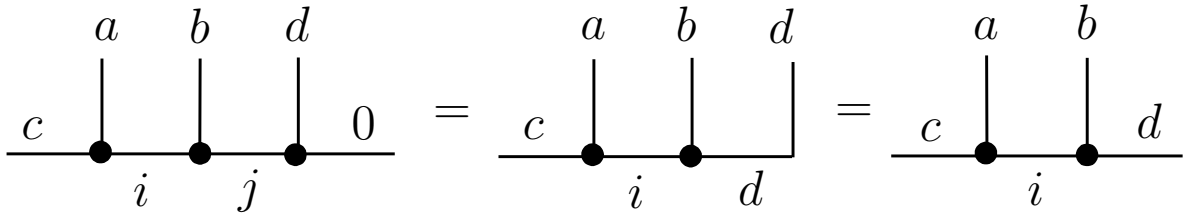}
}
\caption{The configuration of 4 anyons $a$, $b$, $c$, and $d$
fusing into 0 (the vacuum) is equivalent to the configuration of 3
anyons $a$, $b$, and $c$ fusing into one anyon $d$. The first
equality follows from the fact that one can get a 0 only by fusing
an anyon with label $j=d$ with the anyon with label $d$.}
\label{fig:equivalence}
\end{figure}

\begin{figure}
\resizebox{1.00\columnwidth}{!}{%
  \includegraphics{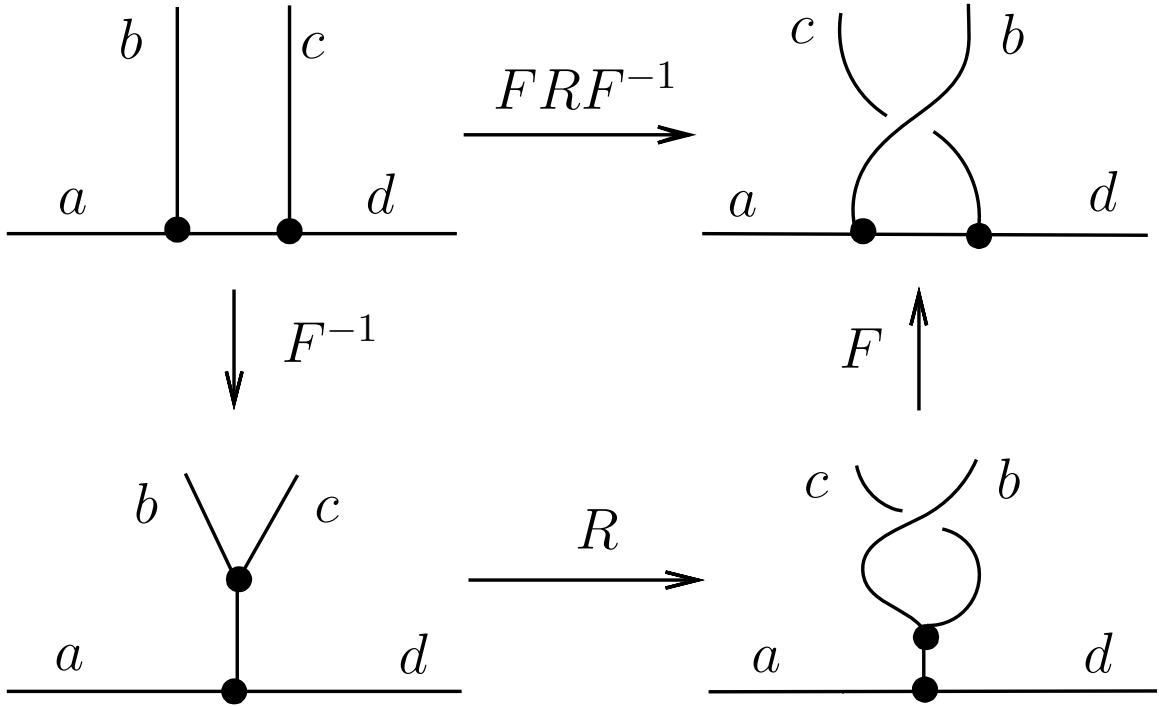}
}
\caption{The use of the F-matrix.}
\label{fig:fmatrix}
\end{figure}

Not all braids are of this case, in which the two braiding anyons
fuse into a single anyon. To see this, it is sufficient to
consider the case of 4 anyons with total spin 0, which is
equivalent to 3 anyons fusing into the 4-th anyon. This is shown
in Fig.~\ref{fig:equivalence}.

\begin{figure}
\resizebox{1.00\columnwidth}{!}{%
  \includegraphics{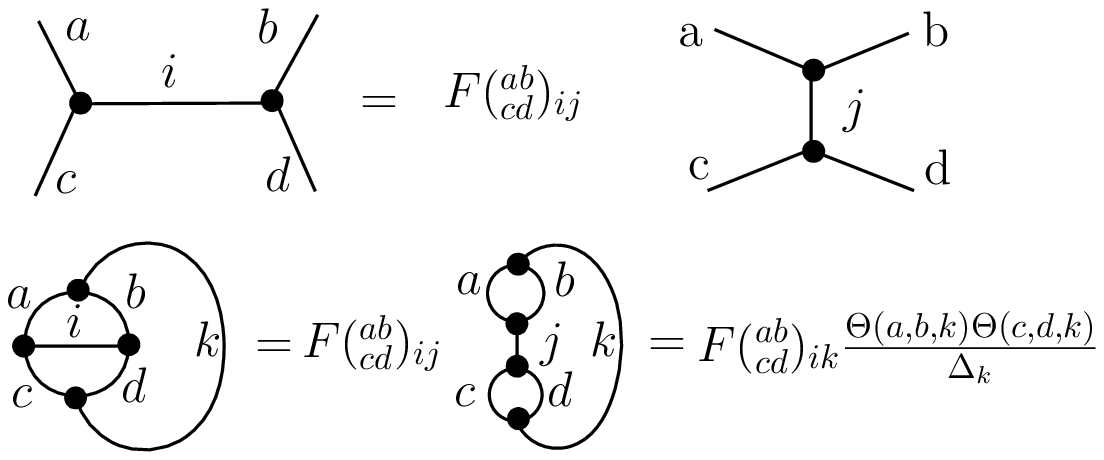}
}
\caption{The upper diagram shows the definition of the matrix
element $F(^{ab}_{cd})_{ij}$ of the F-matrix. A summation on $j$
is implied. The lower diagram shows the derivation of the
expression Eq.~(\ref{eq:oldfmatrix}) of the F-matrix in terms of the delta net, the
theta net, and the tetrahedron net. Expressions in Fig. ~\ref{fig:net} are used
in the derivation.}
\label{fig:fmatrixdef}
\end{figure}

The braiding of $b$ and $c$ in Fig. ~\ref{fig:fmatrix} can not be accomplished via
a single R-matrix and is realized only by a combination of the
R-matrix and the F-matrix. Fig. ~\ref{fig:fmatrixdef}. shows the definition
\cite{kauffman94} of the matrix element $F(^{ab}_{cd})_{ij}$ of
the F-matrix  as well as the formula to calculate it in terms of
the delta net $\Delta_n$, the theta net $\Theta(a,b,c)$, and the
tetrahedron net $T(^{a\;b\;i}_{c\;d\;j})$,
\begin{equation}
F(^{ab}_{cd})_{ij} = \frac{T(^{a\;b\;i}_{c\;d\;j}) \Delta_j}
{\Theta(a,b,j) \Theta(c,d,j)}.
\label{eq:oldfmatrix}
\end{equation}

\subsection{Unitary representations of Artin braid groups}

\begin{figure}
\resizebox{1.00\columnwidth}{!}{%
  \includegraphics{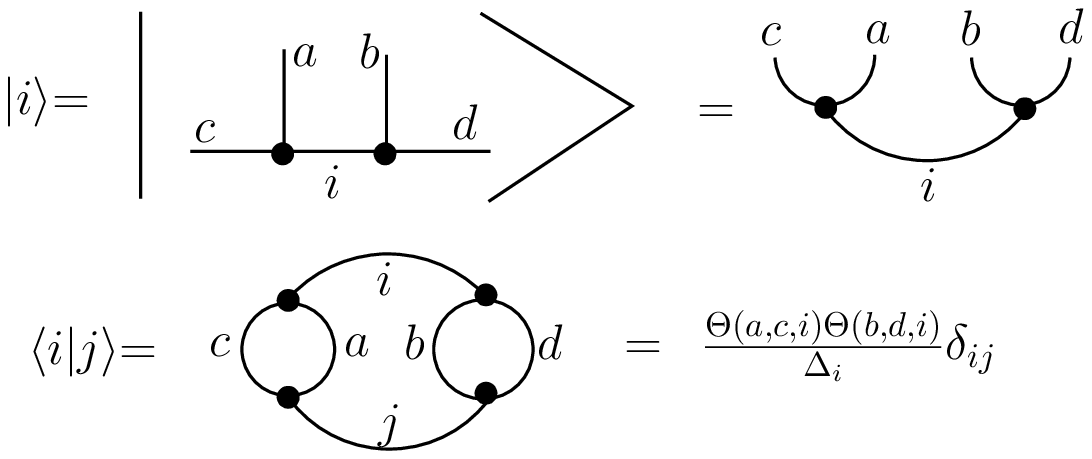}
}
\caption{The upper diagram shows the graphic definition of the
state vector $|i\rangle$ for 4 anyons with total spin 0. The
graphic representation of the dual vector $\langle i|$ is defined
to be the upside-down version of that of the vector $|i\rangle$.
The lower diagram shows the calculation of inner product $\langle
i|j\rangle$ of two states $|i\rangle$ and $|j\rangle$. Expressions
in Fig. ~\ref{fig:net} are needed in the calculation. }
\label{fig:state}
\end{figure}

\begin{figure}
\resizebox{1.00\columnwidth}{!}{%
  \includegraphics{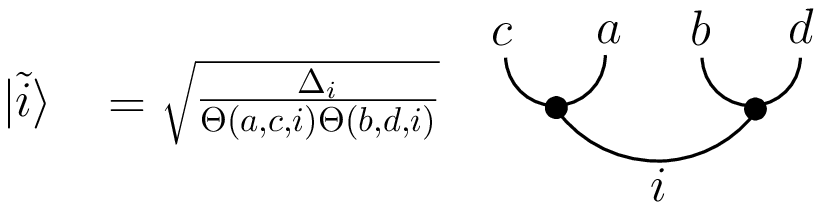}
}
\caption{ The normalized basis $\{|\tilde{i}\rangle\}$ is obtained
from the original one $\{|i\rangle\}$ by multiplying an
appropriate factor to each state $|i\rangle$.}
\label{fig:normalization}
\end{figure}

However, the F-matrix defined above in the Temperley-Lieb
recoupling theory is not unitary, resulting in a non-unitary
representation of the braid group. Unitary (in fact real and
orthogonal) F-matrix, and hence unitary representation of $B_n$
can be obtained by  a redefinition for the basis states in the
Hilbert space of the anyons. The guideline of the following
argument is the requirement that the fusion paths should represent
an orthonormal basis of the Hilbert space of the degenerate ground
states of a system of anyons.

We need only to consider the orthonormal problem of the states of
four anyons fusing into the vacuum. The definition of the state
$|i\rangle$ and the calculation of the inner product $\langle
i|j\rangle$ of the two states $|i\rangle$ and $|j\rangle$ are
shown in Fig. ~\ref{fig:state}. We see that the orthogonal property is already
satisfied ($\langle i|j\rangle\propto\delta_{ij}$), but the state
vectors are not normalized ($\langle i|i\rangle \neq 1$). It follows
that an orthonormal basis $\{|\tilde{i}\rangle\}$ ($\langle
\tilde{i}|\tilde{j}\rangle=\delta_{ij}$) of the Hilbert space is
obtained by normalizing each of the states, as depicted in Fig. ~\ref{fig:normalization}.
In the orthonormal basis $\{|\tilde{i}\rangle\}$, the new F-matrix
can be derived to be
\begin{equation}
F(^{ab}_{cd})_{ij}=\frac{\sqrt{\Delta_i\Delta_j}} {\sqrt
{\Theta(a,b,j)\Theta(c,d,j)\Theta(a,c,i)\Theta(b,d,i)}}
T(^{a\;b\;i}_{c\;d\;j}).
\label{eq:newfmatrix}
\end{equation}
This new F-matrix is real and orthogonal (hence unitary), as we
will see in explicit calculations latter. We note that Kauffman
and Lomonaco \cite{kauffman07} obtained the unitary F-matrix by
multiplying each 3-vertex with the following factor,
\begin{equation}
f(a,b,c)=(\Delta_a\Delta_b\Delta_c)^{1/4}/(\Theta(a,b,c))^{1/2}.
\end{equation}
One can check that this modification to 3-vertices results in the
same F-matrix given by Eq.~(\ref{eq:newfmatrix}).

\begin{figure}
\resizebox{1.00\columnwidth}{!}{%
  \includegraphics{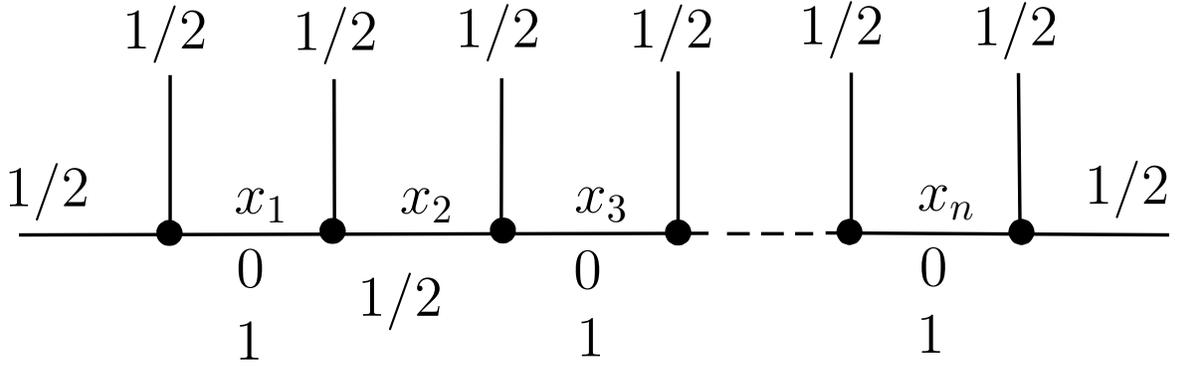}
}
\caption{ The fusion diagram for $n+3$ ($n$ odd) Ising anyons with
total spin 0. The first (the left most one) and the second Ising
anyons fuse into an anyon with spin $x_1=0$ or 1, and then
$x_1$ fuse with the third Ising anyon into an Ising anyon with
spin 1/2, $\cdots$, and finally, $x_n$ fuse with the $(n+2)$-th Ising anyon
into the $(n+3)$-th Ising anyon.}
\label{fig:fusion}
\end{figure}

\section{EBO matrices for Ising anyons}

In this section, we apply the results of section 2 to obtain explicitly the representations of the generators of the braid group governing the exchanges of Ising anyons. Our presentation follows closely to Kauffman  and Lomonaco \cite{kauffman07,kauffman08}.

For the Ising anyons model, the Kauffman variable is
$A=ie^{i\pi/8}$, and the quantum dimension of the spin 1/2 Ising
anyons is $d=-A^2-A^{-2}=\sqrt{2}$.
The allowed spins of anyons in this model are 0, 1/2 and 1 and the
fusion rules for these anyons can be deduced from Eq.~(\ref{eq:fusion}) to be,
\begin{equation}
\begin{array}{ll}
0\otimes j=j \quad \textmd{for} \quad j=0,1/2,1;\\
1/2\otimes1/2=0\oplus 1;\\
1\otimes j=1-j \quad \textmd{for} \quad j=0,1/2,1.\\
\end{array}
\end{equation}
The dimension of the Hilbert space of $n$ spin 1/2 Ising anyons
with total spin 0 is $2^{n/2-1}$, approaching $d^{n}$ in the limit $n\to\infty$.
According to the above fusion rules, the fusion diagram of $n+3$
($n$ must be odd) Ising anyons with total spin 0 takes
the form as shown in Fig.~\ref{fig:fusion} \footnote{For later convenience, we make a change of notation. We will label
anyons by their spins $a/2,b/2,c/2,...$ other than the numbers of
strands $a,b,c,...$ of the corresponding Jones-Wenzl projectors,
which is conventional in the physical literature.}.

We now calculate the unitary representation of $B_{n+3}$ for the braiding of the $n+3$ Ising anyons. We denote the elementary braiding operation (EBO) of the first and the second Ising anyons as $\sigma_1$, the EBO of the second and the third Ising anyons as $\sigma_2$, $\cdots$. The corresponding EBO matrices are denoted by $\rho(\sigma_{i})$ where $1\leq i \leq n+2$.

The first EBO matrix $\rho(\sigma_{1})$ is easy to calculate. It depends only on the label $x_1$. In the basis $\{x_1=|0\rangle,|1\rangle\}$, $\rho(\sigma_{1})$ is simply given by the following R-matrix,
\begin{equation}
R =
\left(\begin{array}{cc} R^{11}_0 & 0\\ 0 & R^{11}_2
\end{array}\right)=
\left(\begin{array}{cc}-A^{-3} & 0 \\ 0 & A\\
\end{array}\right),
\end{equation}
which corresponds to the following Temperley-Leib generator,
\begin{equation}
U = AR - A^2 = \left(\begin{array}{cc} d & 0\\0 & 0
\end{array}\right)
= \left(\begin{array}{cc}
\sqrt{2} & 0\\ 0 & 0\end{array}\right).
\end{equation}
To calculate $\rho(\sigma_{2})$, we need to calculate the following F-matrix in the same basis as above (using Eq.~(\ref{eq:newfmatrix}) and the formulae in the appendix),
\begin{equation}
F = \left(\begin{array}{cc}
F(^{11}_{11})_{00} & F(^{11}_{11})_{02}\\
F(^{11}_{11})_{20} & F(^{11}_{11})_{22}\\
\end{array}\right)\\
=\frac{1}{\sqrt{2}}\left(\begin{array}{cc} 1 & 1 \\ 1 & -1
\end{array}\right),
\end{equation}
corresponding to the following Temperley-Leib generator,
\begin{equation}
V = AS - A^2=
\left(\begin{array}{cc}
1/d & 1/d \\ 1/d & 1/d
\end{array}\right)
=\frac{1}{\sqrt{2}} \left(\begin{array}{cc} 1 & 1 \\ 1 & 1
\end{array}\right),
\end{equation}
where $S=\rho(\sigma_{2})=FRF^{-1}$.

Now consider the case of $\rho(\sigma_{3})$. When either $x_1$ or $x_3$ or both of them equal to 0, the situation is similar to the case of $\rho(\sigma_{1})$ where only R-matrix elements are needed to be calculated. The case in which both $x_1$ and $x_3$ are 1 deserves special consideration. The EBO $\sigma_3$ does not change the value of $x_2$ when both $x_1$ and $x_3$ are 1, and the matrix element of $\rho(\sigma_{3})$ in this case is found to be the same as in the case where both $x_1$ and $x_3$ are 0 by doing some graphic calculations.

The other EBO matrices can be calculated in the same way.
By choosing the basis of the $n+3$ Ising anyons with total spin 0
($n$ odd) as $\{|x_1,x_2,...x_n\rangle\}$ where $x_i$ equals to 0,
1/2, or 1 such that the fusion rules are met at each fusion vertex
along the whole fusion path, as shown in Fig.~\ref{fig:fusion}, we can find a representation of the Temperley-Lieb algebra $TL_{n+3}$. For the first and the last generators of $TL_{n+3}$, we have,
\begin{equation}
\begin{array}{lll}
U_1 |0,x_2,...,x_n\rangle = \sqrt{2}|0,x_2,...,x_n\rangle;\\
U_1 |1,x_2,...,x_n\rangle = 0;\\
U_{n+2} |x_1,...,x_{n-1},0\rangle = \sqrt{2}|x_1,...,x_{n-1},0\rangle;\\
U_{n+2} |x_1,...,x_{n-1},1\rangle = 0.
\end{array}
\end{equation}
For $U_2$ and $U_{n+1}$, we have,
\begin{equation}
\begin{array}{lll}
U_2 |0,1/2,x_3,...,x_n\rangle=\\
\frac{1}{\sqrt{2}} |0,1/2,x_3,...,x_n\rangle +
\frac{1}{\sqrt{2}} |1,1/2,x_3,...,x_n\rangle;\\
U_2 |1,1/2,x_3,...,x_n\rangle=\\
\frac{1}{\sqrt{2}} |0,1/2,x_3,...,x_n\rangle +
\frac{1}{\sqrt{2}} |1,1/2,x_3,...,x_n\rangle;\\
U_{n+1} |x_1,...,x_{n-2},1/2,0\rangle=\\
\frac{1}{\sqrt{2}} |x_1,...,x_{n-2},1/2,0\rangle +
\frac{1}{\sqrt{2}} |x_1,...,x_{n-2},1/2,1\rangle;\\
U_{n+1} |x_1,...,x_{n-2},1/2,1\rangle=\\
\frac{1}{\sqrt{2}} |x_1,...,x_{n-2},1/2,0\rangle +
\frac{1}{\sqrt{2}} |x_1,...,x_{n-2},1/2,1\rangle.
\end{array}
\end{equation}
For the middle ones, we have,
\begin{equation}
\begin{array}{lll}
U_i |x_1,...,x_{i-3},0,1/2,0,x_{i+1},...,x_n\rangle =\\
\sqrt{2} |x_1,...,x_{i-3},0,1/2,0,x_{i+1},...,x_n\rangle;\\
U_i |x_1,...,x_{i-3},0,1/2,1,x_{i+1},...,x_n\rangle = 0;\\
U_i |x_1,...,x_{i-3},1/2,0,1/2,x_{i+1},...,x_n\rangle =\\
\frac{1}{\sqrt{2}} |x_1,...,x_{i-3},1/2,0,1/2,x_{i+1},...,x_n\rangle +\\
\frac{1}{\sqrt{2}} |x_1,...,x_{i-3},1/2,1,1/2,x_{i+1},...,x_n\rangle;\\
U_i |x_1,...,x_{i-3},1/2,1,1/2,x_{i+1},...,x_n\rangle =\\
\frac{1}{\sqrt{2}} |x_1,...,x_{i-3},1/2,0,1/2,x_{i+1},...,x_n\rangle +\\
\frac{1}{\sqrt{2}} |x_1,...,x_{i-3},1/2,1,1/2,x_{i+1},...,x_n\rangle;\\
U_i |x_1,...,x_{i-3},1,1/2,0,x_{i+1},...,x_n\rangle = 0;\\
U_i|x_1,...,x_{i-3},1,1/2,1,x_{i+1},...,x_n\rangle =\\
\sqrt{2} |x_1,...,x_{i-3},1,1/2,1,x_{i+1},...,x_n\rangle.
\end{array}
\end{equation}

The EBO matrices can be obtained immediately from the above representations of the Temperley-Lieb algebra by using Eq.~(\ref{eq:temperleylieb}). One can check that these representations of the braid group indeed satisfy the Artin relations in
Eq.~(\ref{eq:artin}).

\section{Construct quantum gates from the EBO matrices
                for the Ising anyons model}

After obtaining the EBO matrices of the Ising anyons model,
we study in this section some aspects of Ising anyons TQC.

The qubit encoding scheme which is consistent to the quantum circuit model is to use each group of 4 Ising anyons with total spin 0 for each qubit such that an $n$-qubit system uses $4n$ Ising anyons. This is the encoding scheme used by Bravyi \cite{bravyi06} who proved a no-entanglement theorem which states that entangled 2-qubit states can never be prepared by pure topological braiding operations. The proof by Bravyi uses the stabilizer constrains and the no-leakage error conditions. In the following, we give a graphical demonstration of this result from the Temperley-Lieb recoupling approach.

\begin{figure}
\resizebox{1.00\columnwidth}{!}{%
  \includegraphics{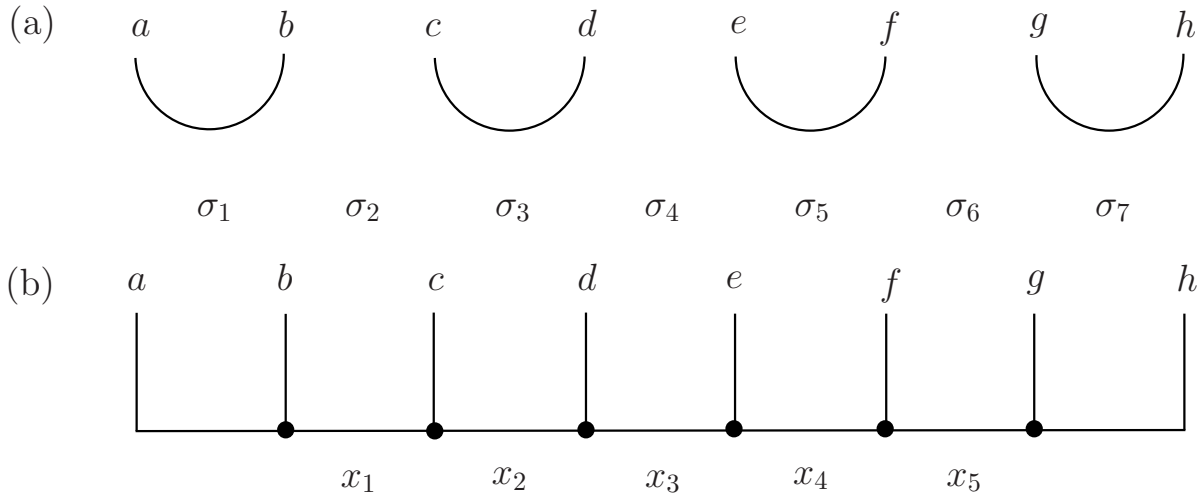}
}
\caption{Part (a) shows the initial state of the two qubits.
The
4 anyons $a$, $b$, $c$, and $d$ form the first qubit and the other
4 anyons $e$, $f$, $g$, and $h$ form the second qubit. Part (b)
shows the full quantum labels needed when we braid the world lines of the
anyons. The computational space is spanned by the states
$\{$
$|x_1,x_5\rangle=$
$|0,0\rangle$,
$|0,1\rangle$,
$|1,0\rangle$,
$|1,1\rangle$
$\}$.
In this notation, the initial state in part (a) is $|0,0\rangle$.
For Ising anyons, $x_2$ and $x_4$ can only be 1/2; for Fibonacci anyons, $x_i$ can be either 0 or 1 such that $x_i+x_{i+1}>0$.}
\label{fig:noentanglement}
\end{figure}

\begin{figure}
\resizebox{1.00\columnwidth}{!}{%
  \includegraphics{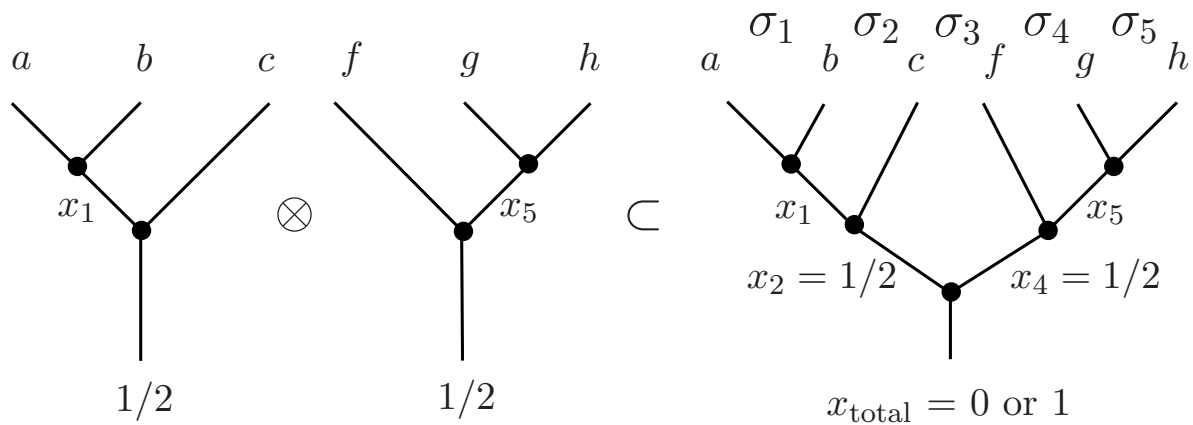}
}
\caption{The formation of the spin 0 ($x_{\rm total}=0$) and spin 1 ($x_{\rm total}=1$) sectors for 8 Ising anyons. Each sector is 4 dimensional and a basis
$\{$
$|x_1,x_5\rangle=$
$|0,0\rangle$,
$|0,1\rangle$,
$|1,0\rangle$,
$|1,1\rangle$
$\}$
can be chosen for both sectors. The rightmost diagram in this figure is related to
part (b) of Fig.~\ref{fig:noentanglement} by a change of basis using
a F-matrix.}
\label{fig:twosectors}
\end{figure}

In this qubit encoding scheme, 2-qubit states are encoded in 8 Ising anyons, 4 Ising anyons for each qubit. See Fig.~\ref{fig:noentanglement}. The two groups of EBOs $\{\sigma_1,\sigma_2\}$ and $\{\sigma_6,\sigma_7\}$ apply completely within the first and the second qubits respectively and can not generate entanglement between the two qubits. Since $\sigma_3$ depends on $x_1$ and $x_3$, $\sigma_5$ depends on $x_3$ and $x_5$, and $\sigma_4$ can change $x_3$ from 0 to a superposition of 0 and 1, it is possible to create an entangled state by a sequence of these EBOs, such as $\sigma_3^{-1}\sigma_4^{-1}\sigma_5\sigma_4\sigma_3$. However, it is impossible to avoid leakage errors by braiding this way. To see this, it is convenient to change the fusion paths in  Fig.~\ref{fig:noentanglement} to another basis as shown in Fig.~\ref{fig:twosectors}, where the total spin of the 2-qubit system can be either 0 or 1. The EBO sequence $\sigma_3^{-1}\sigma_4^{-1}\sigma_5\sigma_4\sigma_3$ in Fig.~\ref{fig:noentanglement} is equivalent to the single EBO $\sigma_3$ in Fig.~\ref{fig:twosectors}. Since the braid matrices of $\sigma_3$ in Fig.~\ref{fig:twosectors} for the two sectors (total spin 0 and 1) are not equivalent, and entanglement can not be created without using $\sigma_3$ in Fig.~\ref{fig:twosectors}, leakage error from the computational space (labeled by $x_1$ and $x_5$ in both Fig.~\ref{fig:noentanglement} and Fig.~\ref{fig:twosectors}) to the uncomputational space (labeled by $x_3$ in Fig.~\ref{fig:noentanglement} and $x_{\rm total}$ in Fig.~\ref{fig:twosectors}) is unavoidable.

It is instructive to compare the above situation with the case of the Fibonacci anyons \cite{preskill219} model which is universal for TQC \cite{freedman02a,freedman02b,freedman03}. See Fig.~\ref{fig:noentanglement} too. For Fibonacci anyons, each intermediate spin $x_i$ can be either 0 or 1 (as long as no two 0s appear consecutively) and entangled states can be generated by braiding the 8 Fibonacci anyons.  Bonesteel \textit{et al} \cite{bonesteel05,simon06,hormozi07} constructed some entangled 2-qubit gates such as the controlled-$iX$ gate by weaving two Fibonacci anyons ($c$ and $d$) from the control qubit into the target qubit which approximates the identity matrix, followed by a braiding within the target qubit which approximates the $iX$ gate, and then weaving them back to their original positions.
The controlling operation is realized by virtue of the fact that the whole braiding does nothing when the total spin of $c$ and $d$ is 0 and acts as the  $iX$ gate on the target qubit when the total spin of $c$ and $d$ is 1. The nonuniversality of the Ising anyons model prevents us from realizing entangled gate in this way.

\begin{figure}
\resizebox{1.00\columnwidth}{!}{%
  \includegraphics{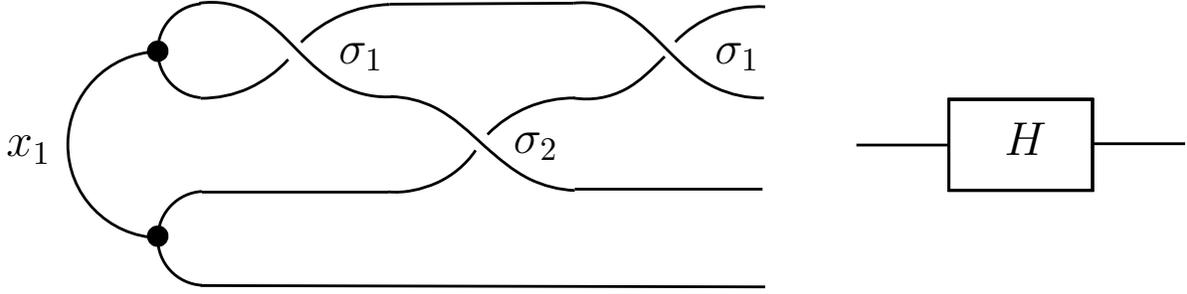}
}
\caption{ The single-qubit states with 4 Ising anyons and the
braiding diagram for the Hadamard gate. $x_1$ = 0 and  $x_1$=1
correspond to the states $|0\rangle$ and $|1\rangle$,
respectively.}
\label{fig:hadamard}
\end{figure}

We note that the no-entanglement theorem only applies to the above qubit encoding scheme where 2 qubits are represented by 8 Ising anyons. Entangled quantum gates can be constructed in a different qubit encoding scheme, as studied by Georgiev \cite{georgiev06,georgiev08a}. In this scheme, 1-qubit and 2-qubit  states are encoded in 4 and 6 Ising anyons with total spin 0 respectively.

Consider 1-qubit gates first.
Taking the basis of the Hilbert space to be (see Fig.~\ref{fig:fusion})
$\{|x_1\rangle=|0\rangle,|1\rangle\}$, the
two dimensional EBO matrices of $B_4$ for four Ising anyons
are found to be
\begin{equation}
\rho^{(2)}(\sigma_1) = \rho^{(2)}(\sigma_3) = e^{i\pi/8}\left(
\begin{array}{cc}
-1 & 0 \\ 0 & i\\
\end{array}\right);
\end{equation}
\begin{equation}
\rho^{(2)}(\sigma_2) = -\frac{e^{-i\pi/8}}{\sqrt{2}}\left(
\begin{array}{cc}
1 & i \\ i & 1
\end{array}\right).
\end{equation}

One can construct the Hadamard gate $H$, the phase
gate $S$, and the three Pauli gates $X$, $Y$, and $Z$ using the two
dimensional EBO matrices given above ($\sim$ means equal up to an
unimportant global phase),
\begin{equation}
H= \frac{1}{\sqrt{2}}\left(
\begin{array}{cc} 1 & 1 \\ 1 & -1\end{array}\right)
\sim \rho^{(2)}(\sigma_1\sigma_2\sigma_1);
\end{equation}
\begin{equation}
S= \left(
\begin{array}{cc} 1 & 0 \\ 0 & i\end{array}\right)
\sim \rho^{(2)}(\sigma_1^{-1});
\end{equation}
\begin{equation}
X= \left(
\begin{array}{cc} 0 & 1 \\ 1 & 0\end{array}\right)
\sim \rho^{(2)}(\sigma_2\sigma_2);
\end{equation}
\begin{equation}
Y= \left(
\begin{array}{cc} 0 & -i \\ i & 0\end{array}\right)
\sim \rho^{(2)}(\sigma_1\sigma_1\sigma_2^{-1}\sigma_2^{-1});
\end{equation}
\begin{equation}
Z=\left(
\begin{array}{cc} 1 & 0 \\ 0 & -1\end{array}\right)
\sim\rho^{(2)}(\sigma_1\sigma_1).
\end{equation}
Fig.~\ref{fig:hadamard} shows the encoding of the 1-qubit states as well as the Hadamard gate constructed by three braids.

However, it fails to construct the $\pi/8$ gate,
\begin{equation}
T=\left(
\begin{array}{cc} 1 & 0 \\ 0 & e^{i\pi/4}\end{array}\right)
\sim \left(\begin{array}{cc} e^{-i\pi/8} & 0 \\ 0 & e^{i\pi/8}
\end{array}\right),
\end{equation}
reflecting the fact that the Ising anyons model is not universal
for quantum computation. To remedy this, we have to supplement braiding with some non-topological operations \cite{bravyi06,freedman06}.

Now consider the 2-qubit case. Taking the basis of the Hilbert space to be (see Fig.~\ref{fig:fusion})
$\{$
$|x_1,x_2,x_3\rangle=$
$|0,1/2,0\rangle$,
$|0,1/2,1\rangle$,
$|1,1/2,0\rangle$,
$|1,1/2,1\rangle$
$\}$,
the four dimensional EBO matrices of $B_6$ for 6 Ising
anyons with total spin 0 read
\begin{equation}
\rho^{(4)}(\sigma_1) =
e^{i\pi/8}\textmd{diag}\left(-1,-1,i,i \right);\\
\end{equation}
\begin{equation}
\rho^{(4)}(\sigma_2) = -\frac{e^{-i\pi/8}}{\sqrt{2}}\left(
\begin{array}{cccc}
1 & 0 & i & 0 \\ 0 & 1 & 0 & i \\ i & 0 & 1& 0 \\ 0 & i & 0 & 1
\end{array}\right);
\end{equation}
\begin{equation}
\rho^{(4)}(\sigma_3)=
e^{i\pi/8}\textmd{diag}\left(-1,i,i,-1\right);
\end{equation}
\begin{equation}
\rho^{(4)}(\sigma_4)=-\frac{e^{-i\pi/8}}{\sqrt{2}}\left(
\begin{array}{cccc}
1 & i & 0 & 0 \\ i & 1 & 0 & 0 \\ 0 & 0 & 1 & i \\ 0 & 0 & i & 1
\end{array}\right);
\end{equation}
\begin{equation}
\rho^{(4)}(\sigma_5) = e^{i\pi/8}\textmd{diag}\left(-1,i,-1,i
\right).
\end{equation}

From the four dimensional EBO matrices given above, one can construct
useful 2-qubit quantum gates, such as CNOT  (up to an
unimportant global phase),
\begin{equation}
\textmd{CNOT} = \left(
\begin{array}{cccc}
1&0&0&0\\0&1&0&0\\0&0&0&1\\0&0&1&0
\end{array}\right)\sim\rho^{(4)}(\sigma_3^{-1}\sigma_4^{-1}
\sigma_5^{-1}\sigma_3 \sigma_4\sigma_3\sigma_1),
\label{eq:cnot}
\end{equation}
and controlled-$Z$ ,
\begin{equation}
\textmd{controlled-$Z$} = \left(
\begin{array}{cccc}
1&0&0&0\\0&1&0&0\\0&0&1&0\\0&0&0&-1
\end{array}\right)\sim\rho^{(4)}(\sigma_1\sigma_3^{-1}
\sigma_5).
\end{equation}

\begin{figure}
\resizebox{1.00\columnwidth}{!}{%
  \includegraphics{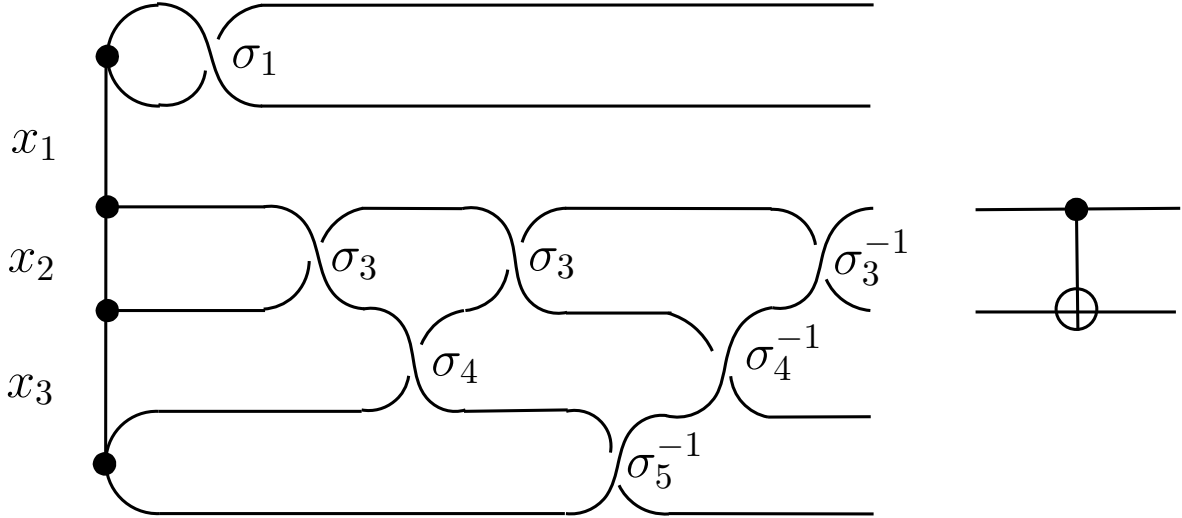}
}
\caption{The 2-qubit states with 6 Ising anyons with total spin 0 and the braiding
diagram for the CNOT gate. $x_1$ and  $x_3$ can be 0 or 1 and
$x_2$ can only be 1/2. Note that the braid sequence for CNOT is not unique, a consequence of the Artin relations for the generators of the braid group.}
\label{fig:cnot}
\end{figure}

Fig.~\ref{fig:cnot} shows the encoding of the 2-qubit states as well as the braiding diagram of CNOT. Note that the braid sequences for CNOT and controlled-$Z$ are not unique, which is  a consequence of the Artin relations Eq.~(\ref{eq:artin}) for the generators of the braid group. Our constructions are different from, but equivalent to the ones given by Georgiev \cite{georgiev06,georgiev08a}.

Note that the 4 dimensional Hilbert space of 6 Ising anyons with total spin 0 is a subspace of the 8 dimensional space of 8 Ising anyons. The other subspace, which is also 4 dimensional, corresponds to 6 anyons with total spin 1. See Fig.~\ref{fig:twosectors}. Using the method in section 3, we can also find the EBO matrices for the spin 1 sector. It turns out that the EBO matrices
$\rho^{(4)}(\sigma_1)$,
$\rho^{(4)}(\sigma_2)$,
$\rho^{(4)}(\sigma_4)$, and
$\rho^{(4)}(\sigma_5)$ in the spin 1 sector take the same form as in the spin 0 sector, but $\rho^{(4)}(\sigma_3)$ has a different form,
$\rho^{(4)}(\sigma_3)= e^{i\pi/8}\textmd{diag}\left(i,-1,-1,i\right)$. Therefore, entangled 2-qubit quantum gates will have different braid sequences in the spin 1 sector. For example, one possible braid sequence in the spin 1 sector for CNOT is $\sigma_5^{-1}\sigma_3^{-1}\sigma_4^{-1}
\sigma_5\sigma_3\sigma_4\sigma_1$,
which is not topologically equivalent to the one given in Eq.~(\ref{eq:cnot}).
Despite of the non-equivalence  of the braid sequences for a given quantum gate in these two sectors, the computational power of the two sectors are equivalent \cite{georgiev08b}.

\section{conclusion and discussion}
As demonstrated in previous sections, the Temperley-Lieb
recoupling theory provides a natural language for describing the
braiding properties of non-Abelian anyons. We have applied this
theory to derive the EBO matrices of the Ising anyons model. We paid a special attention to the normalization of the degenerate ground states corresponding to the fusion paths of the anyons. This normalization results in the correct unitary F-matrices and is equivalent to the redefinition of the 3-vertices proposed by Kauffman and Lomonaco \cite{kauffman07}.

One important feature for the construction of the two-qubit gates is that we can not construct them without the use of $\sigma_3$, the EBO
acting between the two qubits. This is because that the EBOs
$\sigma_1$ and $\sigma_2$ act only on the first qubit and
$\sigma_4$ and $\sigma_5$ act only on the second qubit. Indeed, the first two and the last two EBO matrices can be expressed as a tensor product of two matrices, and the middle EBO matrix $\rho^{(4)}(\sigma_3)$ can not, reflecting the (topological) entanglement of the 2 qubits. This entanglement is crucial for the construction of the 2-qubit entangled gates. However, to get this entanglement, we need to project the $B_8$ representation to either the spin 0 or the spin 1 $B_6$ representations. Alternatively, entangled quantum gates can be constructed by parity measurement as well as braiding operations \cite{bravyi06,zilberberg08}.

The construction of the 2-qubit gates in each sector can be easily achieved by brute force search, since the braid lengths of controlled-$Z$ and CNOT are very short (3 and 7 respectively).  However, there is a more heuristic approach, namely, the genetic algorithm (GA) approach. A possible braid sequence for the CNOT gate can be found within a minute using GA, while it takes a much longer time using the brute force approach. The superiority of GA over brute force search is not significant for Ising anyons TQC, but we expect that there is a potential application of GA to Fibonacci anyons topological quantum compiling \cite{bonesteel05,simon06,hormozi07}.

\section*{Acknowledgements}
We thank Jens~Fjelstad and  Ben~Goertzel for numerous
discussions. We also thank the referees of EPJB who pointed out a mistake of our original manuscript and helped improve this paper a lot.  Z.~Fan is supported by National Natural Science Foundation of China under grant numbers 10535010, 10675090, 10775068, and 10735010.

\appendix
\section{Formulae for evaluating the spin-nets}

In this appendix, we present the formulae for the evaluations of the $\Delta$-net, the $\Theta$-net, and the tetrahedral net \cite{kauffman94}.

The $\Delta$-net evaluation is
\begin{equation}
\Delta_n=(-1)^n[n+1],
\end{equation}
where $[n]$ is the $q$-deformed integer defined as
$[n]=(A^{2n}-A^{-2n})/(A^{2}-A^{-2})$.
The $\Theta$-net evaluation is
\begin{equation}
\Theta(a,b,c)=(-1)^{i+j+k}\frac{[i+j+k+1]![i]![j]![k]!}
{[i+j]![j+k]![k+i]!},
\end{equation}
where the $q$-deformed fractional $[n]!$ is defined as
$[n]!=[n][n-1]...[2][1]$,
and the integers $i$, $j$, and $k$ are determined by the relations
$a=i+j$, $b=j+k$, and $c=k+i$.
The bracket evaluation of the tetrahedral net is
\begin{equation}
\begin{array}{ll}
T(^{a\;b\;i}_{c\;d\;j}) =
\frac{\prod_{m,n}[b_n-a_m]!} {[a]![b]![c]![d]![i]![j]!}
\sum_{\textmd{max}\{a_m\} \leq s \leq \textmd{min}\{b_n\}}
\frac{(-1)^s [s+1]!}{\prod_m [s-a_m]! \prod_n [b_n-s]!},
\end{array}
\end{equation}
where $a_m$  and $b_n$  are given by $a_1=(a+d+i)/2$,
$a_2=(b+c+i)/2$, $a_3=(a+b+j)/2$, $a_4=(c+d+j)/2$,
$b_1=(b+d+i+j)/2$, $b_2=(a+c+i+j)/2$, and $b_3=(a+b+c+d)/2$.

\end{document}